\documentclass[twocolumn,superscriptaddress,nofootinbib]{revtex4}

\usepackage{graphicx,bm,color}

\begin{document}
\title{Dynamic communities in multichannel data: An application to the foreign exchange market during the 2007--2008 credit crisis}

\author{Daniel J. Fenn}
\affiliation{Mathematical and Computational Finance Group, Mathematical Institute, University of Oxford, Oxford OX1 3LB, U.K.}
\affiliation{CABDyN Complexity Centre, University of Oxford, Oxford OX1 1HP, U.K.}
\author{Mason A. Porter}
\affiliation{Oxford Centre for Industrial and Applied Mathematics, Mathematical Institute, University of Oxford, Oxford OX1 3LB, U.K.}
\affiliation{CABDyN Complexity Centre, University of Oxford, Oxford OX1 1HP, U.K.}
\author{Mark McDonald}
\affiliation{FX Research and Trading Group, HSBC Bank, 8 Canada Square, London E14 5HQ, U.K.}
\author{Stacy Williams}
\affiliation{FX Research and Trading Group, HSBC Bank, 8 Canada Square, London E14 5HQ, U.K.}
\author{Neil F. Johnson}
\affiliation{Physics Department, University of Miami, Coral Gables, Florida 33146, U.S.A.}
\affiliation{CABDyN Complexity Centre, University of Oxford, Oxford OX1 1HP, U.K.}
\author{Nick S. Jones}
\affiliation{Physics Department, Clarendon Laboratory, University of Oxford, Oxford OX1 3PU, U.K.}
\affiliation{Oxford Centre for Integrative Systems Biology, University of Oxford, Oxford OX1 3QU, U.K.}
\affiliation{CABDyN Complexity Centre, University of Oxford, Oxford OX1 1HP, U.K.}

\date{\today}

\begin{abstract}
We study the cluster dynamics of multichannel (multivariate) time series by representing their correlations as time-dependent networks and investigating the evolution of network communities. We employ a node-centric approach that allows us to track the effects of the community evolution on the functional roles of individual nodes without having to track entire communities. As an example, we consider a foreign exchange market network in which each node represents an exchange rate and each edge represents a time-dependent correlation between the rates. We study the period 2005-2008, which includes the recent credit and liquidity crisis. Using dynamical community detection, we find that exchange rates that are strongly attached to their community are persistently grouped with the same set of rates, whereas exchange rates that are important for the transfer of information tend to be positioned on the edges of communities. Our analysis successfully uncovers major trading changes that occurred in the market during the credit crisis.
\end{abstract}


\maketitle
\textbf{The last decade has seen a massive explosion of research on static networks and on detecting cohesive subnetworks known as communities. However, despite all of this attention, very little research thus far has considered the problem of investigating communities in time-evolving networks, which occur across disciplines from economics to biology. In this paper, we examine communities in an evolving, weighted, fully-connected network that we obtained from high-frequency time series of currency exchange rates. Community detection in time-dependent weighted networks provides a new approach to the problem of clustering multivariate time series. In contrast to previous studies of dynamic communities in networks, we employ a node-centric approach that allows us to track the effects of the community evolution on the functional role of nodes without having to track entire communities. To illustrate the power of our approach, we consider a foreign exchange (FX) market network during the recent credit crisis. Our analysis uncovers significant structural changes that occurred in the FX market during the crisis and identifies exchange rates that experienced significant, striking changes in market role.}

\section{Introduction}
During the last ten years, there has been an explosion of research on networks (i.e., graphs) with static connections between nodes \cite{NEW_SIAM_2003,ALB_RMP_2002,CALD_BOOK_2007}. This has involved a remarkable interdisciplinary effort -- with physicists, mathematicians, computer scientists, sociologists, and others all playing major roles. Among the most important ideas in network science is the investigation of mesoscopic network structures known as ``communities," \cite{NEW_EPJ_2004,NEWGIV_PRE_2004,DAN_JSM_2005,NEW_PRE_2006,RB_PRE_2006,AREN_ARX_2007,FORT_PNAS_2007,FORT_ARX_2008,FORT_ARX_2009,MAP_SURVEY_2009} which are constructed from subsets of nodes that are more strongly connected to each other than they are to the rest of the network. The analysis of communities has provided striking insights into functional modules in several networks \cite{DAN_JSM_2005,FORT_ARX_2008,FORT_ARX_2009,MAP_PNAS_2005,TRAUD_2009,GUIM_NAT_2005,ADA_BIOINF_2006,MAP_SURVEY_2009}.

However, despite this wealth of attention, there has been much less research that considers communities in the more general class of networks that have time-evolving weighted links \cite{VICSEK_Nat_2007}. Here, we examine the community dynamics in an evolving, fully-connected FX market network over the period 2005-2008 (which includes the recent credit crisis). The network possesses a fixed number of nodes and evolving link weights that are determined by time-varying pairwise correlations between time series associated with each node. Community detection in networks of this kind is equivalent to the problem of clustering multivariate time series. Time series clustering has a range of applications, including finding groups of genes with similar expression patterns in microarray experiments, and grouping functional MRI time series to identify regions of the brain with similar activation patterns \cite{TS_CLUSTERING}. Again, however, there has been little focus on the dynamics of clusters.

In this paper, we present a methodology for investigating cluster/community dynamics. Recently, network scientists have begun to develop techniques to examine the dynamics of communities by tracking entire communities through time \cite{HOPCROFT_2004,FALKOWSKI_2006,VICSEK_Nat_2007}. This requires a method for determining which community at each time step represents the descendant of a community at the previous time step, which can lead to equivocal mappings following splits and mergers. As an alternative, we investigate community dynamics from a \emph{node-centric} perspective and demonstrate how the dynamics can affect the functional role of individual nodes. Our analysis uncovers major changes that occurred in the FX market during the credit crisis, including the identification of individual exchange rates that experienced significant changes in market role. Our approach can potentially provide similarly useful insights into other multivariate data sets from commerce to medicine.


\section{Data}
We construct networks with $n=110$ nodes, where each node represents an exchange rate of the form XXX/YYY (with XXX$\neq$YYY) and XXX, YYY$\in$\{AUD, CAD, CHF, GBP, DEM, JPY, NOK, NZD, SEK, USD, XAU\}. The currency symbols represent: AUD, Australian dollar; CAD, Canadian dollar; CHF, Swiss franc; EUR, euro; GBP, pounds sterling; JPY, Japanese yen; NOK, Norwegian krone; NZD, New Zealand dollar; SEK, Swedish krona; USD, U.S. dollar; XAU, gold. An exchange rate XXX/YYY indicates the amount of currency YYY one can receive in exchange for one unit of XXX. We include gold (XAU) in the study because it behaves similarly to a volatile currency \cite{MM_PRE_2005}.

We define the strength of the link connecting nodes $i$ and $j$ using the time series of hourly exchange rate returns $R_i(t)$ ($i=1,2,\cdots,n$) from the hours 07:00-18:00 U.K. time over the period 2005-2008. The return of an exchange rate with price $p_i(t)$ at discrete time $t$ is defined by $R_i(t)=\ln{\frac{p_i(t)}{p_i(t-1)}}$. We represent the resulting fully-connected, weighted networks by an adjacency matrix $\mathbf{A}$ with components
\begin{equation}
	A_{ij} = \frac{1}{2}(\rho_{ij}+1)-\delta_{ij}\,, \label{network}
\end{equation}
where $\rho_{ij}=\frac{\langle{R_iR_j}\rangle-\langle{R_i}\rangle\langle{R_j}\rangle}{\sigma_i\sigma_j}$ is the correlation coefficient
between exchange rates $i$ and $j$ over a window of $T$ returns, the Kronecker delta $\delta_{ij}$ removes self-edges, $\langle\cdot\rangle$ indicates a time-average over $T$, and $\sigma_i$ is the standard deviation of $R_i$ over $T$. The matrix elements $A_{ij}\in[0,1]$ quantify the similarity of two exchange rates. We use the linear correlation coefficient $\rho_{ij}$ to measure the correlation between pairs of exchange rates because of its simplicity, but our methods are independent of this choice and can be employed using alternative measures capable of detecting more general dependencies.

We exclude self-edges in order to deal with simple graphs. This approach was also taken in a previous study of a financial network derived from a correlation matrix \cite{HEIM_ARX_2008}. However, we note that if we include self-edges, the node compositions of the identified communities are identical if we make a small parameter change in the community detection algorithm. We discuss the community detection algorithm and the effect of including self-edges in Section \ref{sec::comm_detection} and in \cite{self_edges_1,self_edges_2}.

We create a longitudinal sequence of networks by consecutively displacing the time windows by $\Delta{t}=20$ hours (approximately 2 trading days) and fix $T=200$ hours (approximately 1 month of data), as a compromise between over-smoothing and overly-noisy correlation coefficients \cite{JPO_PRE_2003}.

In contrast to most prior studies of financial networks, we use recent community detection techniques \cite{FORT_ARX_2008,FORT_ARX_2009,MAP_SURVEY_2009} rather than traditional hierarchical clustering \cite{MAN_ECON_2000,JPO_PRE_2003,MM_PRE_2005} and consider weighted links rather than thresholding them to create a binary network \cite{FAR_NJP_2007}. In \cite{HEIM_ARX_2008}, the authors also depart from such typical assumptions in a study of communities in a network of equities. However, they do not examine longitudinal networks. Dynamic communities have been investigated in biophysical data using recent methods \cite{SHALIZI_BOOK_2007}, but our work differs by focusing on quantifying and tracking the changes in community composition.


\section{Community detection}
\label{sec::comm_detection}
We represent each network $\mathbf{A}$ as an infinite-range, $n$-state Potts spin glass, so that each node is a spin, each edge is a pairwise interaction between spins, and each community is a spin state. We then determine the community structure by assigning each spin to one state and minimizing the interaction energy of these states, which is given by the Hamiltonian \cite{RB_PRE_2006}
\begin{equation}
	\mathcal{H}=-\sum_{ij}J_{ij}\delta(c_i,c_j),
\label{POTTS}
\end{equation}
where $c_i$ is the state of spin $i$. The interaction energy between spins $i$ and $j$ is $J_{ij}=A_{ij}-{\gamma}p_{ij}$, where $\gamma$ is a resolution parameter and $p_{ij}$ denotes the expected edge weight with which nodes $i$ and $j$ are connected in a null model of random link assignment. The Kronecker delta $\delta(c_i,c_j)$ ensures that the sum is only taken over nodes belonging to the same community. We employ the standard Newman-Girvan null model (or prior), so that $p_{ij}=k_ik_j/2m$, where $k_i=\sum_jA_{ij}$ is the strength (weighted degree) of node $i$ and $m=\frac{1}{2}\sum_{i,j}A_{ij}$ is the total edge weight in the network \cite{NEW_PRE_2006}. For the particular example network considered here, each of the nodes has the same strength $k_i=(n-2)/2$ and the same expected edge weight in the null model given by
\begin{equation}
p_{ij}=\frac{n-2}{2n},
\end{equation}
so the null model is equivalent to the uniform case (with constant $p_{ij}$). However, the methods we present are general and can also be applied to networks with non-uniform strength distributions \cite{self_edges_1}.

By tuning $\gamma$, we probe the community structure at different resolutions, with larger $\gamma$ values corresponding to more fragmented communities \cite{HEIM_ARX_2008,RB_PRE_2006}. At each resolution, we use a greedy algorithm \cite{BLOND_ARX_2008} to minimize Eq.~(\ref{POTTS}). \cite{self_edges_2}

The community structure of many networks is robust across a range of resolutions \cite{RB_PRE_2006,AREN_ARX_2007,FORT_ARX_2008,FORT_ARX_2009}. Robust communities obtained using the Potts method are significant because they persist even though there is a larger incentive for nodes to belong to smaller clusters. We detect communities at 100 resolutions in the interval $\gamma\in[0.8,2.1]$. At $\gamma=0.8$ all of the nodes are assigned to the same community, and at $\gamma=2.1$ they are all assigned to singleton communities. We compare consecutive partitions using the normalized variation of information $\hat{V}$ \cite{MEIL_JMA_2006,TRAUD_2009}. The entropy of a partition $C$ of the $n$ nodes in $\mathbf{A}$ into $K$ communities $c^k$ ($k\in\{1,\cdots,K\}$) is
\begin{equation}
	S(C)=-\sum_{k=1}^{K}P(k)\log P(k),
\end{equation}
where $P(k)={\vert{c^k}\vert}/{n}$ is the probability that a randomly-selected node belongs to community $k$ and $\vert{c^k}\vert$ is the size (set cardinality) of communities. (The quantity $c^k$ is, therefore, the set of nodes labelled by $k$, while $c_i$ is the set of nodes in the same community as node $i$.) Given a second partition $C'$ of the $n$ nodes into $K'$ communities, it follows that
\begin{equation}
	\hat{V}(C,C')=\frac{S(C)+S(C')-2I(C,C')}{\log{n}}.
\end{equation}
The quantity $I(C,C')$, the mutual information between $C$ and $C'$, is given by
\begin{equation}
	I(C,C')=\sum_{k=1}^{K}\sum_{k'=1}^{K'}P(k,k')\log\frac{P(k,k')}{P(k)P(k')},
\end{equation}
where $P(k,k')=\vert{c^{k}\cap{c^{k'}}\vert}/{n}$. The factor $\log{n}$ normalizes $\hat{V}(C,C')$ to the interval $[0,1]$, with $0$ indicating identical partitions and $1$ indicating that all nodes are in individual communities in one partition and in a single community in the other.

\begin{figure}
\includegraphics[width=8.7cm]{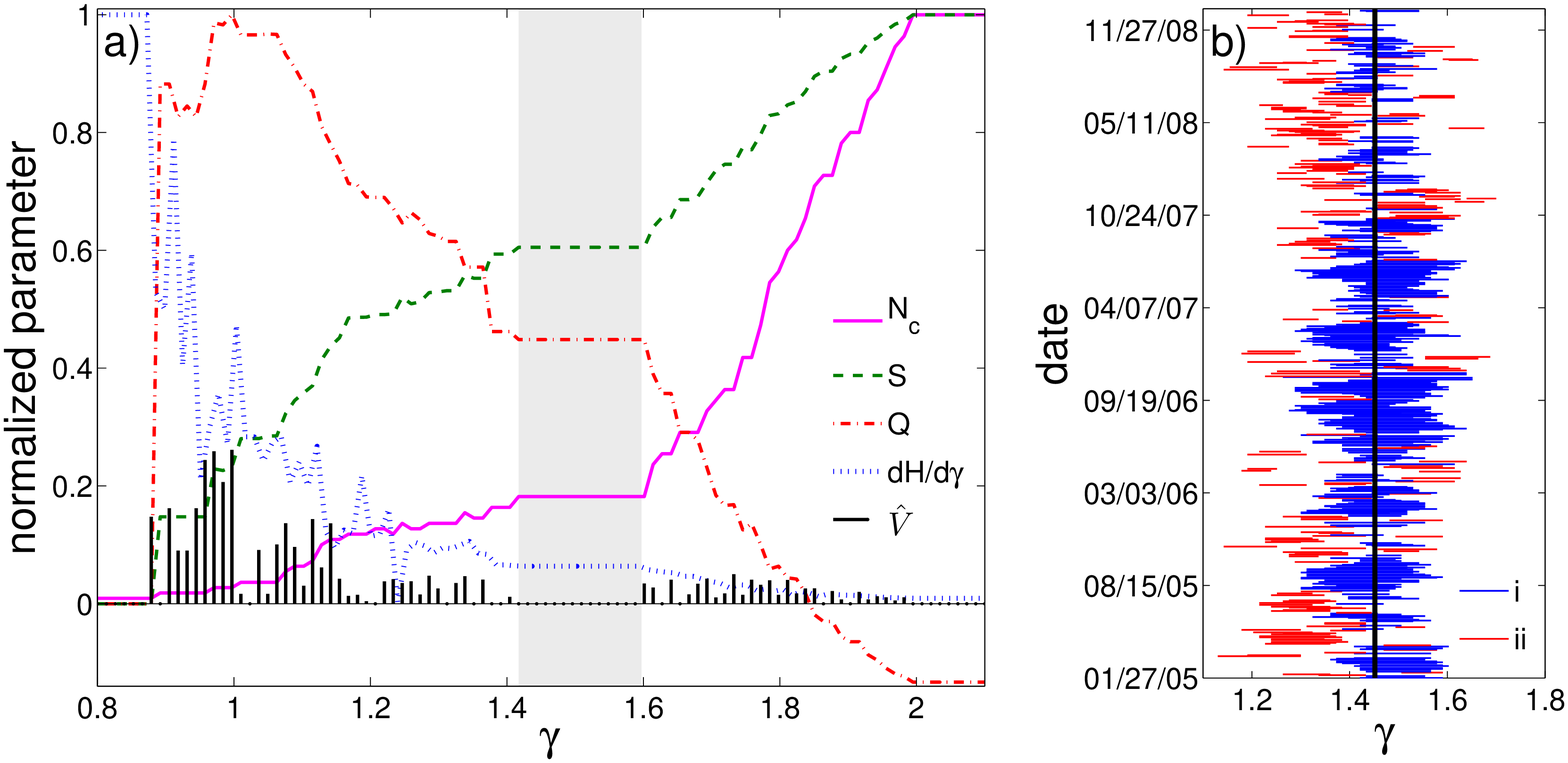}
\caption{(Color online) (a) The quantities $N_c$, $S$, $Q$, and $d\mathcal{H}/d\gamma$ (normalized by their maximum value) as a function of the resolution parameter $\gamma$ for a single time window beginning on 03/10/2005. The shaded gray area highlights the main plateau. The bottom curve gives the normalized variation of information between partitions at resolutions separated by $\Delta\gamma=0.013$. (b) The position of the main plateau at each time-step. Main plateaus (i) containing $\gamma=1.45$ (ii) not containing $\gamma=1.45$.}\label{plateg}
\end{figure}

One can determine robust communities by examining summary statistics that describe community configurations as a function of the resolution parameter. Figure \ref{plateg}(a) shows, as a function of $\gamma$, the number of communities $N_c$, the rate of change of the energy with resolution $d\mathcal{H}/d\gamma$, the modularity $Q$ \cite{NEWGIV_PRE_2004}, and the entropy $S$ of each partition. One can write a scaled energy $Q_s$ in terms of the Hamiltonian in Eq.~(\ref{POTTS}) as $Q_s=-\mathcal{H}/2m$. The modularity $Q$ is the scaled energy with $\gamma=1$.

There are four principle plateaus in Fig. \ref{plateg}(a), corresponding to partitions of the network into $N_c=1$, $2$, $20$, and $110$ communities. The first and last plateaus, respectively, represent all nodes in a single community and all nodes in individual communities. The second plateau represents one community of exchange rates and a corresponding community of inverse rates. We include inverse exchange rates, because one cannot infer \emph{a priori} whether a rate XXX/YYY will form a community with a rate WWW/ZZZ or its inverse ZZZ/WWW. The existence of an equivalent inverse community for each community means that at each time step the network will be formed of two identical halves. The exchange rates residing in each half will, however, change with time as the correlations evolve. For the example in Fig.~\ref{plateg}(a), the other plateau occurs in the interval $\gamma\in[1.41,1.60]$. Although the community configuration over this interval does not have maximal $Q$, it provides an appropriate resolution at which to investigate community dynamics due to its resolution robustness and the financially-interesting features of the detected communities. For the remainder of this paper, we will refer to this plateau as the ``main'' plateau.


\section{Community properties}
\begin{figure}
\includegraphics[width=8.7cm]{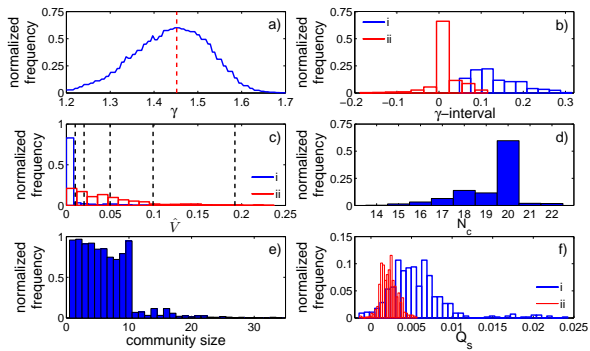}
\caption{(Color online) (a) Observed fraction of time-steps that the resolution $\gamma$ lies on the main plateau. The vertical line indicates the value $\gamma=1.45$, which lies in the highest number of main plateaus and is the resolution at which we investigate the community dynamics. (b) (i) Normalized sampled distribution of the main plateau width and (ii) normalized sampled distribution of the $\gamma$-distance between the main plateau and $\gamma=1.45$. (c) (i) Normalized variation of information distribution between the community configuration at $\gamma=1.45$ and the configuration corresponding to the main plateau and (ii) normalized variation of information distribution between consecutive time steps. The vertical lines give the mean $\hat{V}$ when (left to right) $1$, $2$, $5$, $10$, and $20$ nodes are randomly reassigned to different communities (averaged over 100 reassignments for each time step). (d) The fraction of time-steps at which $N_c$ communities are observed. (e) The fraction of time-steps at which a community of a given size is observed. (f) Comparison of the distribution of the scaled energy for (i) market data and (ii) 100 realizations of shuffled data.}\label{tempstats}
\end{figure}

One way to investigate the community dynamics is to select a resolution $\gamma$ at each time step in the main plateau. As shown in Fig. \ref{plateg}(b), this plateau occurs over different $\gamma$ intervals at different time steps and has different widths. These intervals need not share common resolution values, so this method seems inappropriate because one would then be comparing communities obtained from different resolutions. We therefore fix $\gamma=1.45$, which as shown in Fig. \ref{tempstats}(a) is the value that occurs within the largest number of main plateaus. In order to demonstrate the validity of this technique, we show in Fig. \ref{tempstats}(b) the distribution of the $\gamma$-distance between the fixed resolution $\gamma=1.45$ and the main plateau, and in Fig. \ref{tempstats}(c) the distribution of $\hat{V}$ between the community configuration obtained using $\gamma=1.45$ and that corresponding to the main plateau. The fixed resolution is a $\gamma$-interval of less than 0.05 from the main plateau 88\% of the time, and the community configurations of the main plateau and $\gamma=1.45$ differ in the community assignments of fewer than two nodes 88\% of the time. These results support our proposed method of investigating the community dynamics at a fixed $\gamma$.

Figure \ref{tempstats}(d) shows that there are only small fluctuations in the number of communities into which the network is partitioned. Nevertheless, as shown in Fig. \ref{tempstats}(c), there is a considerable variation in the extent of community reorganization between consecutive time steps. There are no changes between some steps, but more than twenty nodes change communities between others. Figure~\ref{tempstats}(e) shows that the observed community size distribution is bimodal, with a long tail extending to large community sizes. There is then a large variation in the sizes of the communities observed at each time step.

To ensure that the identified communities are meaningful, we compare the scaled energies of the observed community partitions with those from shuffled data. We generate shuffled data by randomly reordering the elements of each of the real time series. By inspection, Fig.~\ref{tempstats}(f) shows that the communities identified for the actual data are significantly stronger than those generated using shuffled data. The sample mean scaled energy for the actual data is $0.0056$ (with a standard deviation of $0.0036$) and for shuffled data the sample mean is $0.0022$ (with a standard deviation of $0.0010$).


\section{Node roles in communities}
Having considered the properties of entire communities, we now investigate the roles of nodes within communities. A node's identity is known at all times and its community is known at any one time. We can thus track community evolution from the perspective of individual nodes. We describe the relationship between a node and its community using various centrality measures. The betweenness centrality, commonly studied by sociologists, is defined as the number of geodesic paths between pairs of vertices in a network \cite{FREE_SOCI_1977,NEW_SIAM_2003}. We take the distance between nodes $i$ and $j$ as $d_{ij}=1/A_{ij}$ for $i\neq{j}$ and $0$ for $i=j$. The betweenness centrality, in some sense, measures the importance of each node for the spread of information around the network.

We also consider the community centrality of each node \cite{NEW_PRE_2006}. We define a scaled energy matrix $\mathbf{J}$ by $J_{ij}=A_{ij}-\gamma{p_{ij}}$, where we again set $p_{ij}=k_ik_j/2m$. Following the notation in \cite{NEW_PRE_2006}, this matrix can be expressed as $\mathbf{J}=\mathbf{UDU}^T$, where $\mathbf{U}=(\mathbf{{u_1}\vert{u_2}\vert\cdots})$ is the matrix of eigenvectors of $\mathbf{J}$ and $\mathbf{D}$ is the diagonal matrix of eigenvalues $\beta_i$. If $\mathbf{D}$ has $q$ positive eigenvalues, one can define a set of node vectors $\mathbf{x}_i$ by
\begin{equation}
	[\mathbf{x}_i]_j=\sqrt{\beta_j}U_{ij},\qquad{j\in\{1,2,...,q\}}.
\end{equation}
The magnitude $\vert\mathbf{x}_i\vert$ is the \textit{community centrality}. Nodes with high community centrality play an important role in their local neighborhood irrespective of the community boundaries.

One can also define a community vector
\begin{equation}
	\mathbf{w}_k=\sum_{i\in{c}^k}\mathbf{x}_i.
\end{equation}
for each community $k$ with members $c^k$. Nodes with high community centrality are strongly attached to their community if their node vector is also aligned with their community vector. We then define a \textit{projected community centrality} $y_i$ by
\begin{equation}
	y_i=\mathbf{x}_i\cdot\hat{\mathbf{w}}_k=\vert\mathbf{x}_i\vert\cos{\theta_{ik}},
\end{equation}
where $\hat{\mathbf{w}}_k$  is the unit vector in the direction of $\mathbf{w}_k$, and refer to the quantity $\cos{\theta_{ik}}$ as the \emph{community alignment}. The community alignment is near $1$ when a node is at the center of its community and near $0$ when it is on the periphery. Nodes with high community centrality, which are located in the center of their community, have a high projected community centrality and hence are attached strongly to their community (and can be considered to be highly influential within it). We normalize $\vert\mathbf{x}_i\vert$ and $y_i$ by the maximum value at each time step.

We investigate the persistence through time of nodes' communities by defining a community autocorrelation. For a node $i$ with community $c_i(t)$ at time $t$, the autocorrelation $a_i^t(\tau)$ of its community after $\tau$ time steps is defined by
\begin{equation}
	a_i^t(\tau)=\frac{\mid{c}_i(t)\cap{c}_i(t+\tau)\mid}{\mid{c}_i(t)\cup{c}_i(t+\tau)\mid}.
\end{equation}
This is a node-centric version of a quantity considered in \cite{VICSEK_Nat_2007}. Importantly, this measure does not require one to determine which community at each time step represents the descendant of a community at the previous time step. Instead, the communities are identified from the perspective of individual nodes.

\begin{figure}
\includegraphics[width=8.7cm]{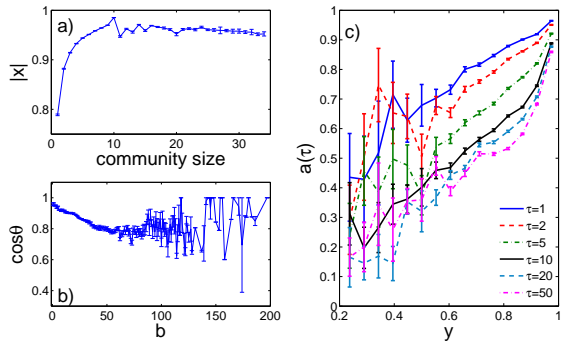}
\caption{(Color online) (a) Mean community centrality versus the size of the community to which the node belongs. (b) Mean community alignment versus the betweenness centrality of nodes. (c) Mean community autocorrelation versus the projected community centrality. (All error bars indicate the standard error \cite{STATS_BOOK}.)}\label{station}
\end{figure}

In Fig. \ref{station}(a) we show the mean normalized community centrality as a function of community size (averaging over all nodes belonging to the same size community). The community centrality increases with community size up to sizes of about $10$ members. For larger communities, $\vert\mathbf{x}_i\vert$ remains approximately constant. Nodes with high $\vert\mathbf{x}_i\vert$ therefore tend to belong to large communities, so exchange rates with high community centrality tend to be closely linked with many other rates. Figure \ref{station}(b) shows the mean betweenness centrality versus the community alignment (averaging the community alignment over all nodes with the same betweenness). Nodes with high betweenness centrality tend to have small values for their community alignment, implying that nodes that are important for the transfer of information are usually located on the edges of communities.

In Fig. \ref{station}(c), we show the mean community autocorrelation versus the projected community centrality. We calculate the mean autocorrelation by splitting the range of $y$ into $15$ equally spaced bins and then averaging over all autocorrelations falling within these bins. (The observed relationships are robust for reasonable variations in the number of bins.) As one would expect, the community autocorrelation for the same projected community centrality is smaller for larger $\tau$. For all values of $\tau$, the mean community autocorrelation increases with $y$. This suggests that nodes that are strongly connected to their community are likely to persistently share that community membership with the same subset of nodes. In contrast, exchange rates with a low $y$ experience regular changes in the set of rates with which they are clustered.


\section{Effects of the credit crisis on network structure}
\begin{figure}
\includegraphics[width=8.7cm]{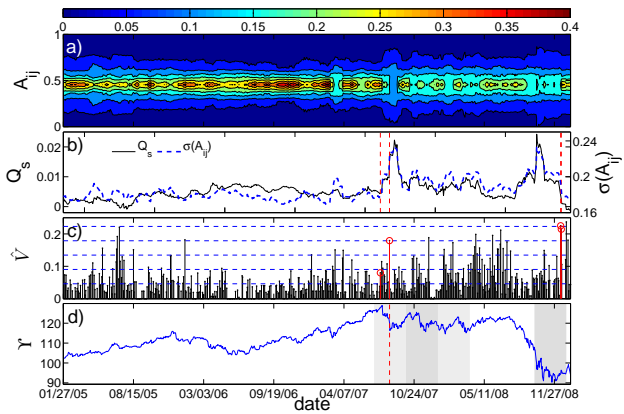}
\caption{(Color online)
(a) Normalized distribution of link weights at each time-step. (b) Scaled energy $Q_s$ and standard deviation of the link weights. (c) Normalized variation of information between the community configuration at consecutive time steps. The horizontal lines show (from bottom to top) the mean of $\hat{V}$ and $1$, $2$, $3$, and $4$ standard deviations above the mean. The vertical lines in (b) and (c) show the time steps when the following dates enter the rolling time window: 07/20/07, 08/15/07, and 12/16/08 (two large reorganizations at consecutive time steps are marked following this date). (d) Carry trade index $\Upsilon$. The vertical line shows 08/15/07 and the shaded blocks (from left to right) Q3 2007, Q4 2007, Q1 2008, and Q4 2008.}\label{evolution}
\end{figure}

Thus far, we have considered the community properties aggregated over all time steps. We now investigate the community dynamics over specific time intervals. In particular, we focus on the insights that such shorter-term dynamics can provide into the changes that occurred in the FX market during the recent credit crisis. Figure \ref{evolution}(a) shows a contour plot of the normalized distribution of link weights at each time-step. The mean link strength remains constant through time because of the inclusion in the network of each exchange rate and its inverse but [as one can see in Figs. \ref{evolution}(a,b)] there is a large variation in the standard deviation of the link strengths. Figure \ref{evolution}(b) shows that the scaled energy is closely related to the standard deviation of the link weights. This is expected because the standard deviation increases as a result of the strengthening of strong ties and the weakening of weak ties.

In Fig. \ref{evolution}(c), we also show $\hat{V}$ between the community configuration at consecutive time steps. Significant changes in the configuration are indicated by large spikes in $\hat{V}$. The correlation coefficient between $\hat{V}$ and the absolute change in $Q_s$ between consecutive time steps is 0.47, and that between $\hat{V}$ and the absolute change in $\sigma(A_{ij})$ is 0.27. Changes in $\sigma(A_{ij})$ are therefore not always a good indicator that there has been a change in the community configuration of the network.

The seriousness of the credit crisis first became widely recognized when on 07/19/07 Federal Reserve chairman Ben Bernanke warned in testimony to the U.S. Congress that the crisis in the U.S. sub-prime lending market could cost up to \$100 billion. This announcement marks the start of a prolonged increase in $Q_s$ beginning as 07/20/07 enters the rolling time window. There is no significant community reorganization at this time step.

The most important effect of the credit crisis on the FX market was its impact on the carry trade. The first significant community reorganization of the crisis occurred on 08/15/07 as a result of massive ``unwinding'' of the carry trade. This change accompanied a large increase in $Q_s$. The carry trade consists of selling low interest rate ``funding currencies'' such as the JPY and CHF and investing in high interest rate ``investment currencies'' such as the AUD and NZD. It yields a profit if the interest rate differential between the funding and investment currencies is not offset by a commensurate depreciation of the investment currency \cite{BRUNNERMEIER_2008}. The carry trade requires a strong appetite for risk, so unwinding of the trade tends to occur during periods in which there is a decrease in available credit. A trader unwinds a carry trade position by selling their holdings in investment currencies and buying funding currencies.

One approach to quantifying carry trade activity is to consider the returns that could have been achieved using a carry trading strategy. The total return for a currency trading strategy can be split into two parts: a return resulting from the price changes and a return resulting from the interest rate differentials between the currencies. We can estimate the return that could have been achieved for a particular carry trade strategy using historical price and interest rate time series. We consider a common strategy in which one buys equal weights of the three major currencies with the highest interest rates and sells equal weights of the three currencies with the lowest interest rates. In Fig. \ref{evolution}(d), we show the cumulative return index $\Upsilon$ for this trading strategy. Large negative returns result in large decreases in $\Upsilon$, which are therefore likely to indicate significant unwinding of the carry trade.

\begin{figure*}
\includegraphics[width=12.5cm]{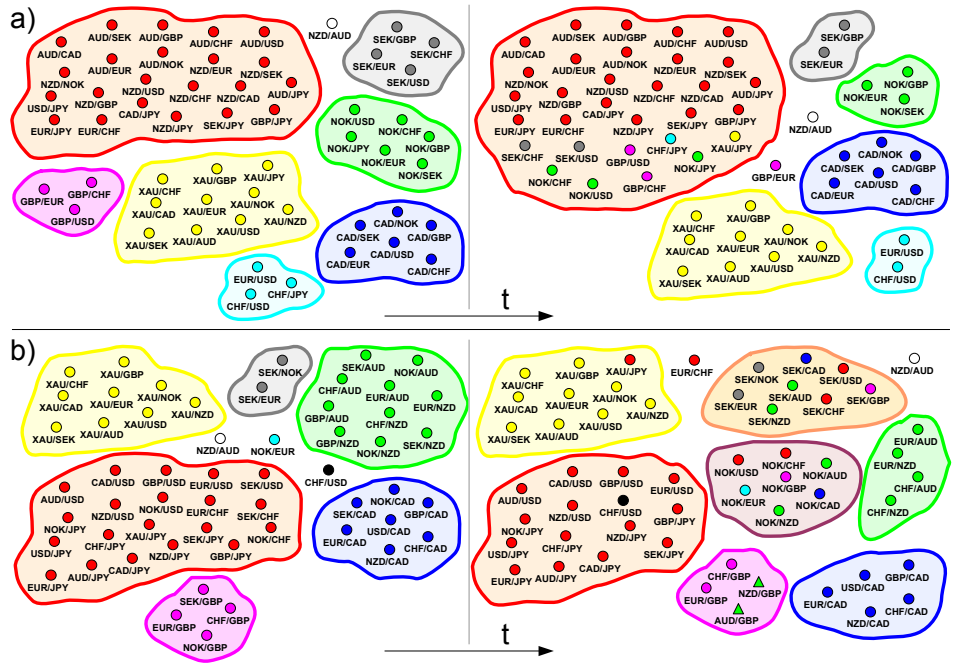}
\caption{(Color online)
Schematic representation of the change in the community configuration in one half of the FX market network following: (a) 08/15/07, when there was significant unwinding of the carry trade (b) 12/16/08, when the U.S. Federal Reserve cut the funding interest rate from $1\%$ to $0-0.25\%$. The node colors after the community reorganization correspond to their community before the change. If the parent community of a community after the reorganization is obvious it is drawn in the same color as its parent. The nodes represented as triangles resided in the opposite half of the network before the community reorganization.}\label{comm_schematic}
\end{figure*}

In Fig. \ref{comm_schematic}(a) we show the observed communities before and after 08/15/07. Figure \ref{evolution}(d) shows that leading up to 08/15/07 there has been some unwinding of the carry trade, so the initial configuration includes a community containing exchange rates of the form AUD/YYY, NZD/YYY, and XXX/JPY (which all involve one of the key carry trade currencies). During 08/15/07-08/17/07 there is a sharp increase in carry unwinding resulting in this community increasing in size as it incorporates other XXX/JPY rates as well as some XXX/CHF and XXX/USD rates. The number of exchange rates involving one of the key carry trade currencies in a single community clearly demonstrates the significance of the trade over this period. Some of the exchange rates included in the community are also somewhat surprising and provide insights into the range of currencies used in the carry trade.

In Fig.~\ref{comm_schematic}(b), we show an example of a significant community reorganization that accompanied a decrease in $Q_s$. This reorganization occurred following 12/16/08, when the U.S. Federal Reserve cut the federal funds interest rate from $1\%$ to $0-0.25\%$ -- the lowest rate ever recorded.

We have considered three examples in which large community reorganizations are clearly attributable to particular market events, demonstrating the effectiveness of our methods. However, there are several other large community changes that are not easily associated with specific events. Investigating the community changes at these time-steps might provide insights into market changes that are not otherwise obvious.

Finally, we consider the change in the relationship between specific nodes and their communities during the crisis period. We begin by defining within-community $z$-scores, which directly compare the relative importance of different nodes to their community \cite{GUIM_NAT_2005}. We describe the roles of individual nodes at each time step using the within-module projected community centrality $z$-score $z^y$ and within-module betweenness centrality $z$-score $z^b$. If a node $i$ belongs to community $c_i$ and has projected community centrality $y_i$, then $z^{y}_i=(y_i-\bar{y}_{c_i})/\sigma^{y}_{c_i}$, where $\bar{y}_{c_i}$ is the average of $y$ over all nodes in $c_i$ and $\sigma^{y}_{c_i}$ is the standard deviation of $y$ in $c_i$. The quantity $z^{y}_i$ measures how strongly connected node $i$ is to its community. Similarly, if the same node has betweenness centrality $b_i$, then $z^b_i=(b_i-\bar{b}_{c_i})/\sigma^b_{c_i}$, where $\bar{b}_{c_i}$ is the average of $b$ over all nodes in $c_i$ and $\sigma^{b}_{c_i}$ is the standard deviation of $b$ in $c_i$. The quantity $z^b_i$ indicates the importance of node $i$ to the spread of information compared with other nodes in the same community. The positions of nodes in the $(z^b,z^y)$ plane thereby illuminate the roles of the associated exchange rates in the FX market.

\begin{figure}
\includegraphics[width=8.7cm]{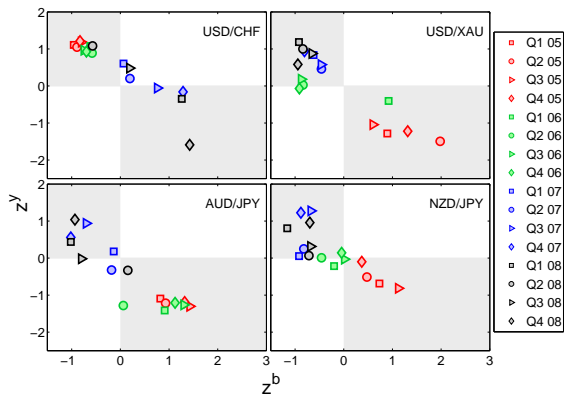}
\caption{(Color online) Quarterly node role evolutions in the $(z^b,z^y)$ plane.}\label{roles}
\end{figure}

We investigate the time dynamics of exchange rate roles by examining quarterly changes in their positions in the $(z^b,z^y)$ plane. The quarterly roles are determined by averaging $z^y$ and $z^b$ over all time steps in each quarter. Changes in a node's position in the $(z^b,z^y)$ plane reflect changes in the membership of a node's community as well as changes in $b$ and $y$.

We show four example role evolutions in Fig.~\ref{roles}. The USD/XAU rate provides an interesting example due to the persistence of its community. From 2005-2008, the USD/XAU node has shifted from being an important information carrier within the XAU community to being more central to this community. This period of higher influence coincides closely with the period of financial turmoil. The CHF is widely regarded as a ``safe haven'' currency \cite{RANALDO_2008}, so one might expect USD/CHF to behave in a similar manner to USD/XAU. However, the CHF is also a key carry trade currency. The use of the CHF as both a safe haven and carry trade currency means that USD/CHF does not move in the same direction as USD/XAU in the $(z^b,z^y)$ plane. Instead the USD/CHF exchange rate is an important information carrier during the credit crisis. The AUD/JPY and NZD/JPY exchange rates change from being important for information transfer to being influential within their communities. These exchange rates are the ones that are most widely used for the carry trade, so their increased importance is unsurprising. In fact, both rates were highly influential during the third and fourth quarters in 2007, the first quarter in 2008, and the fourth quarter of 2008, when there was significant carry trade activity [see Fig. \ref{evolution}(d)].


\section{Conclusions}
In summary, we have analysed the evolving community structure in time-dependent networks to provide new insights into the clustering dynamics of multichannel/multivariate time series. We focused, in particular, on a node-centric community analysis that allows one to follow the time dynamics of the functional role of individual nodes within networks. As an illustration of our approach, we examined the FX market network during a period that included the 2007-2008 credit crisis. We demonstrated that the FX market has a robust community structure over a range of resolutions and showed that there is a relationship between an exchange rate's functional role and its position within its community. Our analysis of the community dynamics successfully uncovered significant structural changes that occurred in the FX market during the credit crisis and identified exchange rates that experienced significant changes in market role. Our methodology should be similarly insightful for other multivariate data sets.


\begin{acknowledgments}
We thank S.~D. Howison, A.~C.~F. Lewis, M.~A. Little P.~J. Mucha, J.-P. Onnela, S. Reid, and S.~J. Roberts and his group for suggestions and code. We acknowledge HSBC bank for providing the data. N.~.S.~J. acknowledges the BBSRC and EPSRC. M.~A.~P. acknowledges a research award (\#220020177) from the James S. McDonnell Foundation.
\end{acknowledgments}



\begin{thebibliography}{10}
\bibitem{self_edges_1}
If we include self-edges, the strength of each node will increase by one. This, in turn, leads to a constant increase in the expected edge weight in the null model. For a network with self-edges, the expected edge weight is given by $p^{s}_{ij}=n/[2(n+2)]$, a shift by a constant value of $p_{ij}^s-p_{ij}= 2/[n(n+2)]=1.62\times{10^{-4}}$ relative to the network in which self-edges are excluded. Self-edges always occur within a community, so they will always contribute to the summation in Eq.~(\ref{POTTS}) irrespective of exactly how the nodes are partitioned into communities. This implies that self-edges play no role when selecting the community partition that minimizes the interaction energy at a particular resolution.
\bibitem{self_edges_2}
In order to find equivalent communities in the network in which self-edges are included, it is necessary to decrease the resolution parameter to compensate for the increase in the constant expected edge weight in the null model. If we identify communities in the network in which self-edges are excluded using the resolution parameter $\gamma$, then we find identical communities in the corresponding network with self-edges using a resolution parameter $\gamma^{s}=\gamma{p_{ij}}/{p^{s}_{ij}}=\gamma(n+2)(n-2)/n^2$. For example, if we identify communities in the network without self-edges using a resolution of $\gamma=1.4500$, then we identify equivalent communities in the network with self-edges with a resolution parameter of $\gamma_s=1.4495$.
\end{thebibliography}
\end{document}